\documentclass[aps,prb,amsmath,amssymb,reprint,superscriptaddress]{revtex4-2}

\usepackage{graphicx}
\usepackage{amsmath,amssymb}
\usepackage{xcolor}
\usepackage{color,soul}
\renewcommand\hl[1]{#1} 

\begin{document}

\title{Inverse Sandwich Geometry and Stability of B$_7$Y$_2$ Clusters: A DFT Study}

\author{P. L. Rodr\'iguez-Kessler}
\email{plkessler@cio.mx}
\affiliation{Centro de Investigaciones en \'Optica A.C., Loma del Bosque 115, Lomas del Campestre, Leon, 37150, Guanajuato, Mexico}

\date{\today}

\begin{abstract}
In this work, we use density functional theory (DFT) to investigate the structural and electronic properties of B$_7$Y$_2$ clusters—boron frameworks doped with two yttrium atoms. Our results show that the most stable configuration adopts an inverse sandwich geometry, while higher-energy isomers ($\sim$1.45 eV above) exhibit B$_7$ wheel-like motifs with yttrium at top or bridge sites. To understand the bonding and stability, we analyze the electron localization function (ELF) and localized orbital locator (LOL) maps. The global minimum shows a symmetric, delocalized electron distribution, strong B–B covalent bonding, and partial charge transfer from Y atoms. In contrast, higher-energy isomers show less effective Y–B interactions. ELF and LOL analyses confirm electron delocalization within the boron framework. Mulliken population analysis reveals significant metal-to-boron charge transfer, contributing to the stability of the inverse sandwich structure through synergistic effects of delocalized bonding and charge redistribution.

\end{abstract}


\maketitle

\section{Introduction}

Boron clusters continue to captivate researchers due to their rich structural diversity and unique bonding characteristics, which differ significantly from those of carbon-based systems. The addition of transition or rare-earth metals to boron clusters has opened new avenues in the design of novel nanomaterials with tailored properties. In particular, doped boron clusters have exhibited remarkable electronic and structural behaviors, such as planar aromaticity and metal-centered coordination geometries.\cite{Zhai2003-fu} Among these, inverse sandwich structures—where a metal atom or atoms are encapsulated between boron rings—have attracted considerable interest for their stability and potential applications in materials science.\cite{doi:10.1021/acs.accounts.4c00380,doi:10.1126/science.196.4294.1047}

Doping boron clusters with other elements can significantly modify their electronic, magnetic, and chemical properties, thereby expanding their potential for diverse applications.\cite{D4CP00296B} Notably, studies involving single transition metal (TM) dopants have led to the discovery of a wide range of doped boron cluster geometries, including wheel-like,\cite{D0NJ03999C} tubular,\cite{doi:10.1073/pnas.0408132102} cage-shaped,\cite{C5NR04034E} umbrella-like,\cite{doi:10.1021/acs.jpca.6b12232} and bilayer structures.\cite{D0NR09214B} Representative examples include Ta@B$_{20}$,\cite{C6CC09570D} the half-sandwich neodymium-doped B$_{12}^-$ cluster,\cite{C9CP03496J} and B$_{16}$ and B$_{24}$ clusters doped with a variety of TMs such as V, Cr, Mn, Fe, Co, and Ni.\cite{C9NJ06335H}

\indent While studies on single-doped boron clusters are well-established, double-doped boron clusters remain less explored.\cite{Chen_2025,doi:10.1021/acs.inorgchem.7b02585,molecules28124721,RODRIGUEZKESSLER2024122062,RODRIGUEZKESSLER2023116538} A pioneering investigation by Zhai et al. in 2006 examined the geometric and bonding characteristics of the binary Au$_2$B$_7^-$ cluster.\cite{doi:10.1021/jp0559074} Later, in 2014, Li and co-workers conducted combined experimental and theoretical analyses of Ta$_2$B$_6^{-/0}$ clusters, identifying D$_{6h}$ bipyramidal structures as the most stable forms for both neutral and anionic species.\cite{https://doi.org/10.1002/anie.201309469} More recently, Wang and colleagues examined lanthanide-doped boron clusters, Ln$_2$B$_n$ (n = 7, 9), and discovered the presence of inverse sandwich motifs within this series.\cite{doi:10.1073/pnas.1806476115} In particular, studies on M$_2$-doped boron clusters M$_{2}$B$_7$ involving alkaline earth metals (M = Be, Mg, Ca) as well as first-row transition metals found typical inverse sandwich structures.\cite{JIA2014128,Zhuan-Yu2014,PHAM2019186,C5CP01650A,LI202325821,OLALDELOPEZ2024,RODRIGUEZKESSLER2025117486} 

Despite the growing interest in boron-containing compounds, only a limited number of studies have focused on Y–B clusters. Yttrium–boron systems are of particular interest due to their potential applications in materials science, including electronics, magnetism, and the design of novel functional materials.\cite{PhysRevB.76.214103} Recently, Guevara et al. investigated the structures of monodoped Y$_1$B$_n$ clusters and evaluated their bonding and aromatic characteristics.\cite{https://doi.org/10.1002/cphc.202400544} To further explore these intriguing materials, we focus in this study on B$_7$Y$_2$ clusters—a boron framework doped with two yttrium atoms—to investigate their structural preferences using density functional theory (DFT). Although several bonding motifs are conceivable for this system, our comprehensive structural search identifies the inverse sandwich configuration as the global minimum. This geometry is energetically favored by a significant margin over alternative isomers, which generally exhibit Y$_2$-doped B$_7$ wheel-like arrangements. To further characterize these structures, we analyze their electron localization functions (ELFs) and localized orbital locator (LOL) maps, offering deeper insights into their bonding nature and overall stability.

\section{Computational Details}

Calculations performed in this work are carried out by using density functional theory (DFT) as implemented in the ORCA 6.0.0 code.\cite{10.1063/5.0004608} The exchange and correlation energies are addressed by the PBE0 functional in conjunction with the Def2-TZVP basis set.\cite{10.1063/1.478522,B508541A} Atomic positions are self-consistently relaxed through a Quasi-Newton method employing the BFGS algorithm. The SCF convergence criterion is set to TightSCF in the input file. This results in geometry optimization settings of 1.0e$^{-08}$ Eh for total energy change and 2.5e$^{-11}$ Eh for the one-electron integrals. The  Van  der  Waals  interactions  are  included in the exchange-correlation functionals with empirical dispersion corrections of Grimme DFT-D3(BJ). The electron localization function (ELF) was computed and analyzed using Multiwfn.\cite{https://doi.org/10.1002/jcc.22885} 

\section{Results}

The most stable structures of B$_7$Y$_2$ clusters are identified using a modified basin-hopping (MBH) structure search method, following approaches described in previous works.\cite{D0CP06179D,al12m,D4CP04444D} Eight initial structures are randomly perturbed multiple times and subjected to the Metropolis criterion until 100 structures are selected. These structures are then evaluated through single-point energy calculations and ranked accordingly. The lowest-energy configurations are fully optimized and used to seed the next generation. The MBH method enhances standard random perturbations by incorporating random exchanges of atomic species, enabling a more efficient exploration of the potential energy surface of binary alloys.

\begin{figure}[h!]
  \begin{tabular}{c}
      \includegraphics[scale=0.7]{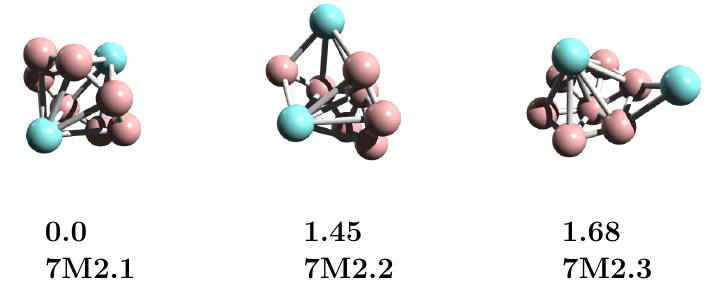} \\
      \end{tabular}
       \caption{\label{figure_struc}Lowest energy structures for B$_7$Y$_{2}$ clusters. The clusters are labeled by the {\bf 7M2.y} notation.\cite{b7cr2clusters,D4CP04444D,b7al2,al12m,pt5v} For each structure, the relative energy (in eV) and isomer label are given. }
\end{figure}

The results show that the lowest energy structure for B$_7$Y$_{2}$ cluster is an inverse sandwich (IS) structure ({\bf 7M2.1}) with spin multiplicity two (m = 2). From the structural point of view, the IS structure prevails for a number of B$_7$M$_{2}$ (M = transition metal) clusters.\cite{ZHANG2015131,HAO20181,JIA2014128} The next isomer adopt a distorted structure with Y$_2$ at the peripheral site ({\bf 7M2.2}) and relative energy ($\Delta$E) of \hl{1.45} eV (Figure~\ref{figure_struc}), while the 3rd isomer ({\bf 7M2.3}) adopt a Y$_2$ supported B$_7$ wheel structure, with $\Delta$E = 1.68 eV.

\begin{figure}[ht]
\caption{\label{figure_2}The electron localization function (ELF = 0.8) for a) {\bf 7M2.1} and b) {\bf 7M2.2} clusters.}
 \resizebox*{0.40\textwidth}{!}{\includegraphics{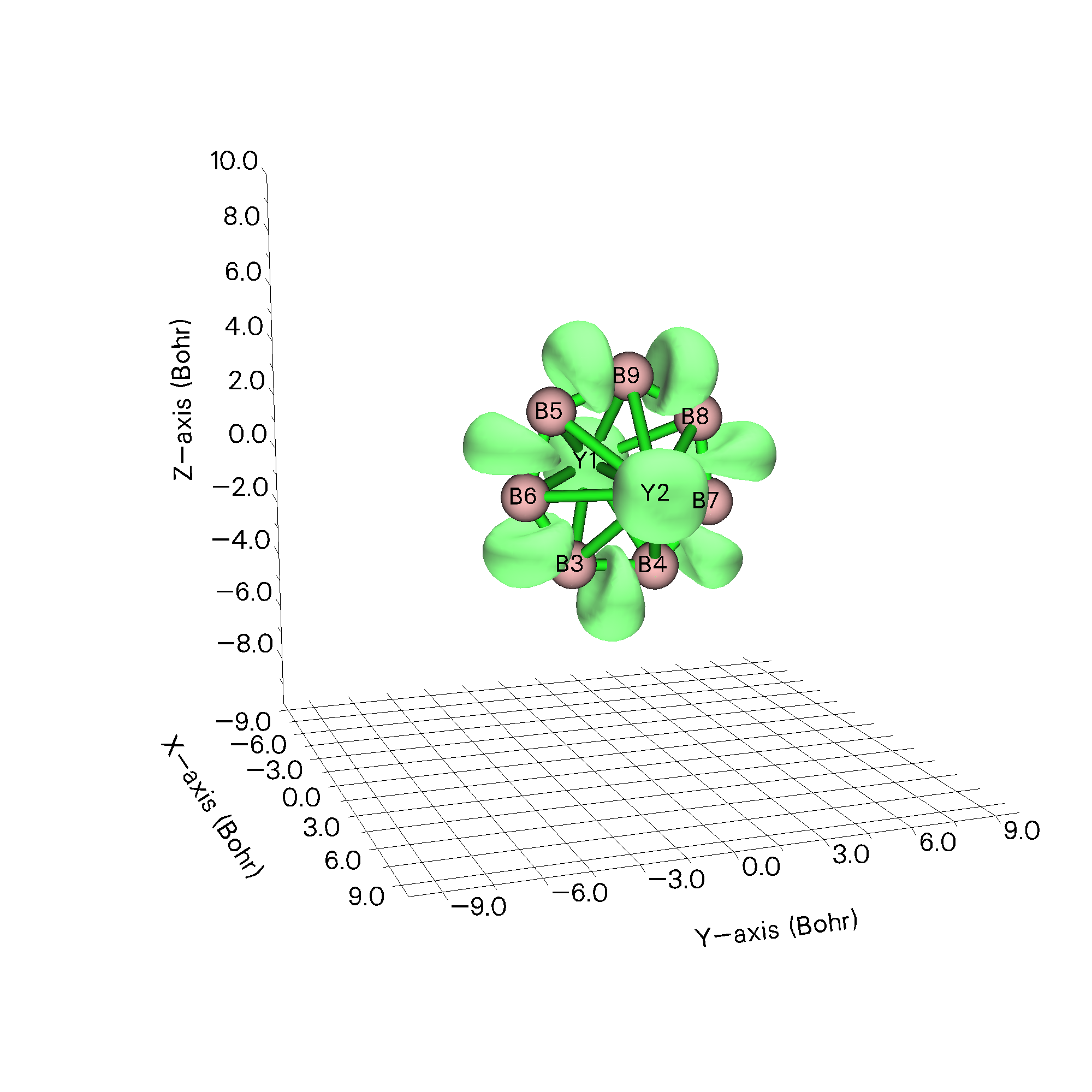}} 
\resizebox*{0.40\textwidth}{!}{\includegraphics{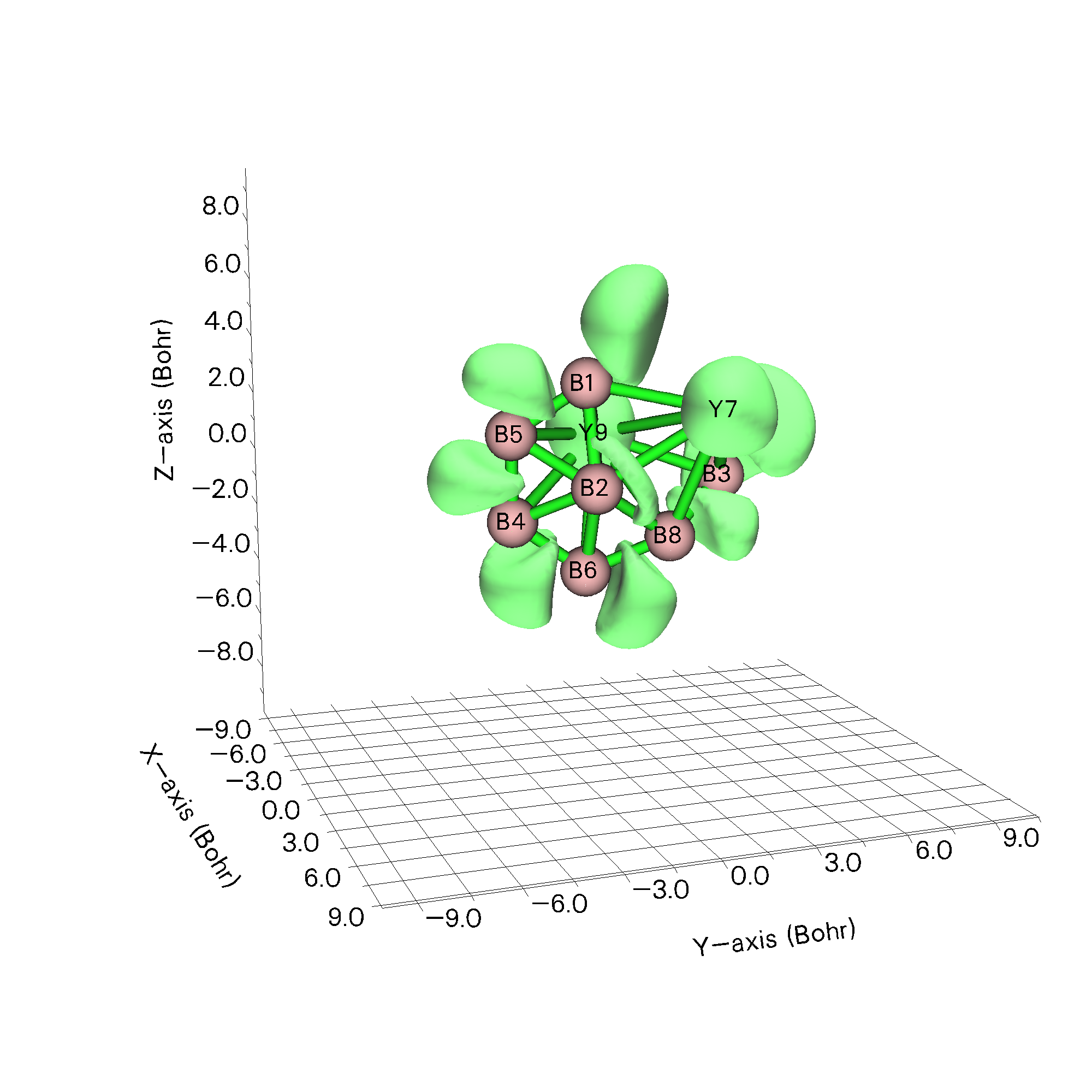}} 
\end{figure}

\begin{table}[ht!]
{
\caption{\label{table_1}{Mulliken population analysis of the most stable B7Y2 cluster. }}
\small
\def\arraystretch{1.1}
\begin{tabular}{p{1.6cm}p{1.6cm}p{1.6cm}p{1.6cm}p{1.6cm}}
\hline
   Atom   &   Alpha pop. &  Beta pop.  &  Spin pop.  &   Atomic charge \\
     1(Y )   &   5.36314   &   5.04955  &   0.31358    &    28.58731\\
     2(Y )   &   5.36303   &   5.04940  &    0.31363   &     28.58757\\
     3(B )   &   2.67534   &   2.53226  &    0.14308   &     -0.20761\\
     4(B )   &   2.63165   &   2.54670  &    0.08494   &     -0.17835\\
     5(B )   &   2.73442   &   2.44839  &    0.28603   &     -0.18281\\
     6(B )   &   2.50044   &   2.62769  &   -0.12725   &     -0.12814\\
     7(B )   &   2.52575   &   2.60497  &   -0.07923   &     -0.13072\\
     8(B )   &   2.74239   &   2.47659  &    0.26579   &     -0.21898\\
     9(B )   &   2.46385   &   2.66442  &   -0.20057   &     -0.12827\\
\hline
\end{tabular}
	}
\end{table}

\begin{figure*}[htb!]
{
\caption{\label{fig_3}{Localized orbital locator (LOL) function for the global minimum structure of B$_7$Y$_2$ cluster.}}
\small
\def\arraystretch{1.1}
\begin{tabular}{cc}
\resizebox*{0.30\textwidth}{!}{\includegraphics{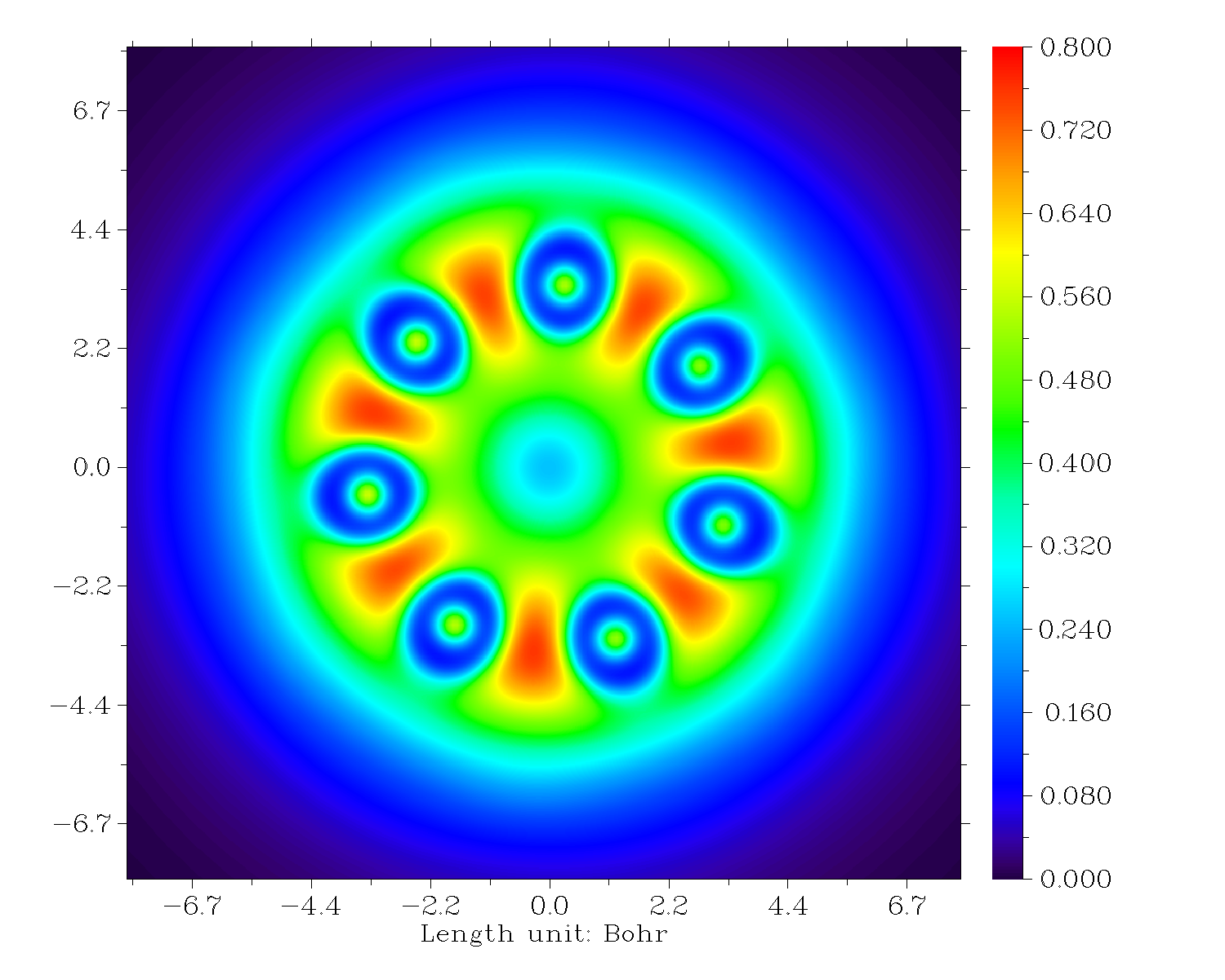}} &
\resizebox*{0.30\textwidth}{!}{\includegraphics{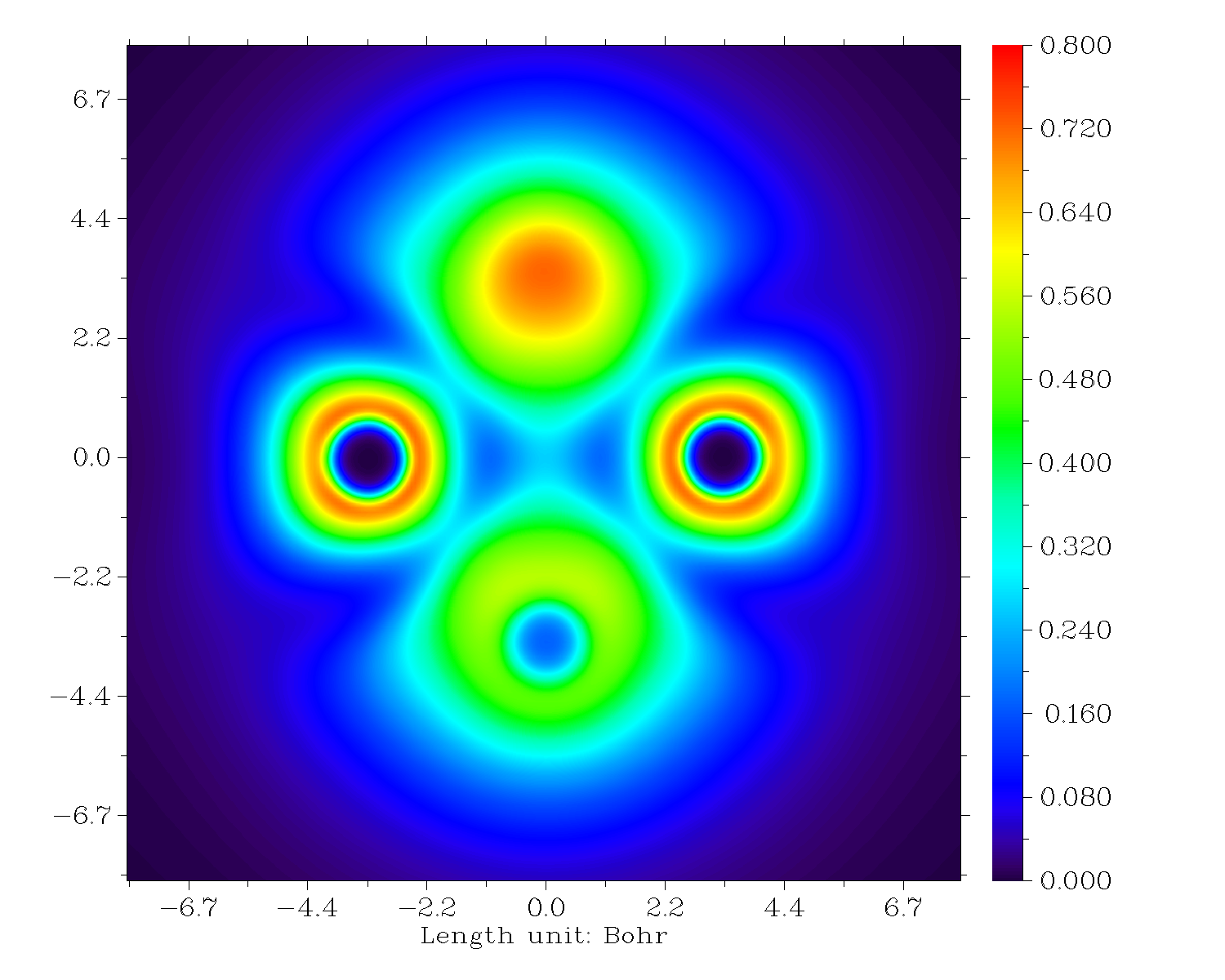}} \\
\end{tabular}
	}
\end{figure*}

To gain insight into the bonding stability of the clusters, we evaluated the electron localization function (ELF) analysis of the clusters, which, provides valuable insights into the electronic structure and stability of the isomers,\cite{C7CP01740E,MAI201987,2025stabreactdoubleicosahedron} as depicted in Figure~\ref{figure_2}. In the most stable configuration (upper panel), the ELF reveals a symmetric and delocalized electron distribution among the boron atoms, indicative of strong covalent B–B interactions and effective electron sharing with the yttrium atoms. The green ELF lobes between B–B atoms clearly represent localized bonding regions, while the presence of ELF lobes around the Y atoms indicates areas of polarized electron density, suggesting partial charge transfer or interaction between B and Y atoms. The symmetric arrangement of these lobes contributes to a well-organized electronic structure and supports the stability of the cluster. In contrast, the low-lying isomer (lower panel) displays a more distorted ELF pattern, with less uniform electron localization and altered Y–B interactions. The ELF lobes around the Y atoms in this structure are more isolated and asymmetrically distributed, implying less effective bonding and reduced electron delocalization compared to the most stable isomer. These differences in ELF topology correlate well with the relative energies of the isomers, highlighting the critical role of electron localization and metal-boron coordination in stabilizing B$_7$Y$_2$ clusters.

The Mulliken population analysis of the B$_7$Y$_2$ cluster reveals important details about charge distribution and spin polarization within the system (see Table~\ref{table_1}). The two yttrium atoms exhibit nearly identical electron populations, with slight spin polarization (spin populations ~0.31), and significant positive atomic charges of approximately +28.59, indicating that they donate electron density to the boron framework and act as electron-deficient centers. In contrast, the boron atoms carry small negative charges (ranging from approximately –0.13 to –0.22), suggesting that the boron network acts as the primary electron acceptor. This is consistent with the expected electron-rich nature of boron in metal-doped boron clusters. The spin population is unevenly distributed among the boron atoms, with certain atoms (such as B$_5$ and B$_8$) showing notable spin densities ($\sim$0.26–0.29), while others (like B$_6$, B$_7$, and B$_9$) exhibit negative spin populations, indicating spin polarization and possible delocalization of the unpaired electron across the boron ring. Overall, the Mulliken analysis supports the idea of a delocalized electronic structure with significant charge transfer from the yttrium atoms to the boron cluster, contributing to the stability and unique bonding features of the B$_7$Y$_2$ system. These features confirm that the electronic structure of the most stable B$_7$Y$_2$ cluster is governed by a delocalized boron framework stabilized through charge transfer from the metal atoms.

Moreover, Figure~\ref{fig_3} presents the localized orbital locator (LOL) function calculated via density functional theory (DFT) for the most stable Y$_2$B$_7$ cluster, shown from front (left) and side (right) perspectives. The LOL function provides a spatial visualization of electron localization, which is valuable for understanding the bonding characteristics within the cluster. In both views, regions of high electron localization are indicated by red and yellow zones, primarily surrounding the boron framework, reflecting strong covalent interactions between boron atoms. The central and ring-like distributions in the front view suggest delocalized $\pi$-type bonding typical of boron-rich clusters. Meanwhile, the side view highlights the out-of-plane electron density, pointing to three-dimensional bonding interactions and possible electron delocalization over the entire cluster. The presence of yttrium atoms subtly perturbs the electron distribution, likely contributing to the cluster’s structural stability by donating electron density or stabilizing specific orbitals. Overall, the LOL maps indicate a complex bonding network dominated by delocalized and covalent B–B interactions, stabilized by the presence of the yttrium atoms. These results provide a comprehensive understanding of the bonding and stability mechanisms in B$_7$Y$_2$ clusters, offering valuable insights for the design of novel boron-based materials and transition metal-doped nanostructures.\cite{RODRIGUEZKESSLER2025122376}

\section{Conclusions}
In summary, our DFT investigation of the B$_7$Y$_2$ cluster reveals that the most energetically favorable structure is an inverse sandwich geometry, stabilized by a combination of strong covalent B–B bonding and effective electron donation from the yttrium atoms. Analyses of the ELF and LOL functions demonstrate a highly delocalized and symmetric electron distribution within the boron framework, characteristic of aromatic-like bonding, while also highlighting the role of yttrium in modulating the electronic structure. The comparison with higher-energy isomers underscores the importance of electron localization and metal–boron coordination in determining cluster stability. Mulliken population analysis further supports the significant charge transfer and spin polarization associated with the yttrium dopants. 


\section{Acknowledgments}
P.L.R.-K. would like to thank the support of CIMAT Supercomputing Laboratories of Guanajuato and Puerto Interior. 



\bibliographystyle{unsrt}
\bibliography{mendelei.bib}
\end{document}